\documentclass[twocolumn,10pt,superscriptaddress,prl]{revtex4}
\usepackage{ifthen}
\usepackage{color}
\usepackage{graphics,graphicx,epsfig}
\usepackage{epsf,epstopdf,wrapfig}
\usepackage{amssymb,amsfonts,amsmath}
\usepackage[export]{adjustbox}
\include{epsf}
\usepackage{ifthen}

\newcommand{\beqn}{\begin{eqnarray}}
\newcommand{\eeqn}{\end{eqnarray}}
\newcommand{\beq}{\begin{equation}}
\newcommand{\eeq}{\end{equation}}

\definecolor{junglegreen}{rgb}{0.16, 0.67, 0.53}
\definecolor{myrtle}{rgb}{0.13, 0.26, 0.12}
\definecolor{lincolngreen}{rgb}{0.11, 0.35, 0.02}
\definecolor{forestgreen}{rgb}{0.13, 0.55, 0.13}

\begin{document}
\title{SOS: Online probability estimation and generation of T and B cell receptors}

\author{Giulio Isacchini}
\affiliation{Laboratoire de physique de l'\'Ecole normale sup\'erieure
  (PSL University), CNRS, Sorbonne Universit\'e, and Universit\'e de
  Paris, 75005 Paris, France}
\affiliation{Max Planck Institute for Dynamics and Self-organization, Am Fa\ss berg 17, 37077 G\"ottingen, Germany}
\author{Carlos Olivares}
\affiliation{Laboratoire de physique de l'\'Ecole normale sup\'erieure
  (PSL University), CNRS, Sorbonne Universit\'e, and Universit\'e de
  Paris, 75005 Paris, France}
\author{Armita Nourmohammad}
\affiliation{Max Planck Institute for Dynamics and Self-organization, Am Fa\ss berg 17, 37077 G\"ottingen, Germany}
\affiliation{Department of Physics, University of Washington, 3910
  15th Avenue Northeast, Seattle, WA 98195, USA}
\affiliation{Fred Hutchinson cancer Research Center, 1100 Fairview ave N, Seattle, WA 98109, USA}
\author{Aleksandra M. Walczak}
\thanks{These authors contributed equally.}
\affiliation{Laboratoire de physique de l'\'Ecole normale sup\'erieure (PSL University), CNRS, Sorbonne Universit\'e, and Universit\'e de Paris, 75005 Paris, France}
\author{Thierry Mora}
\thanks{These authors contributed equally.}
\affiliation{Laboratoire de physique de l'\'Ecole normale sup\'erieure
  (PSL University), CNRS, Sorbonne Universit\'e, and Universit\'e de
  Paris, 75005 Paris, France}

\begin{abstract}
Recent advances in modelling VDJ recombination and subsequent selection of T and B cell receptors provide useful tools to analyze and compare immune repertoires across time, individuals, and tissues.
A suite of tools---IGoR~\cite{Marcou2018}, OLGA~\cite{Sethna2019} and SONIA~\cite{Sethna2020}---have been publicly released to the community that allow for the inference of generative and selection models  from high-throughput sequencing data. 
However using these tools requires some scripting or command-line skills and familiarity with complex datasets. As a result the application of the above models has not been available to a broad audience.
In this application note we fill this gap by presenting Simple OLGA \& SONIA (SOS), a web-based interface where users with no coding skills can compute the generation and post-selection probabilities of their sequences, as well as generate batches of synthetic sequences. The application also functions on mobile phones.\\
{\bf Availability and implementation:} SOS is freely available to use at \url{sites.google.com/view/statbiophysens/sos} with source code at \url{github.com/statbiophys/sos}.
\end{abstract}

\maketitle

\section{Introduction}
The adaptive immune systems recognises pathogens through the
generation of a highly diverse repertoire of T and B cell receptors
(TCR and BCR) which have the potential to recognise even unknown
pathogen and initiate an immune response. In order to produce this
diversity it exploits a highly stochastic process named V(D)J
recombination. In addition, to block possible auto-reactive receptors,
a selection process is mounted in the thymus for T cells, and a
similar process of central tolerance is implemented for B
cells. Probabilistic models of TCR and BCR have
been proposed~\cite{Murugan2012,Ralph2016,Elhanati2014} based on  immune
repertoire sequencing 
data~\cite{Georgiou2014,Heather2017,Minervina2019a,Bradley2019}. Software
has been developped to infer the probability of generation
of any B- or T-cell receptor (IGoR~\cite{Marcou2018}), and to evaluate
this probability for both nucleotide and amino-acid
sequences (OLGA~\cite{Sethna2019}). Another tool
(SONIA~\cite{Sethna2020}) was released to infer the
selective pressures acting on the receptors and used to predict the probability of naive sequences in the periphery \cite{Isacchini2020}. In order to make these tools available to a broader audience, we provide a new web tool which allows for the analysis of single TCR and BCR sequences.

\section{Features}
As explained in the introductory ``About'' tab, the web tool evaluates
the generation and post-selection probability of single naive T and B cell receptors in different species based on the specific sequence the user inputs manually. 
The engine is based on two pieces of python software, OLGA and SONIA,
and shipped with pre-trained models of recombination and
selection for the following loci: human alpha and beta
chains or TCR (TRA and TRB), human heavy and light chain of unmutated BCR (IGH,
IGK, and IGL), and mouse TRB.

\begin{figure*}
\begin{center}
\includegraphics[width=.65\linewidth]{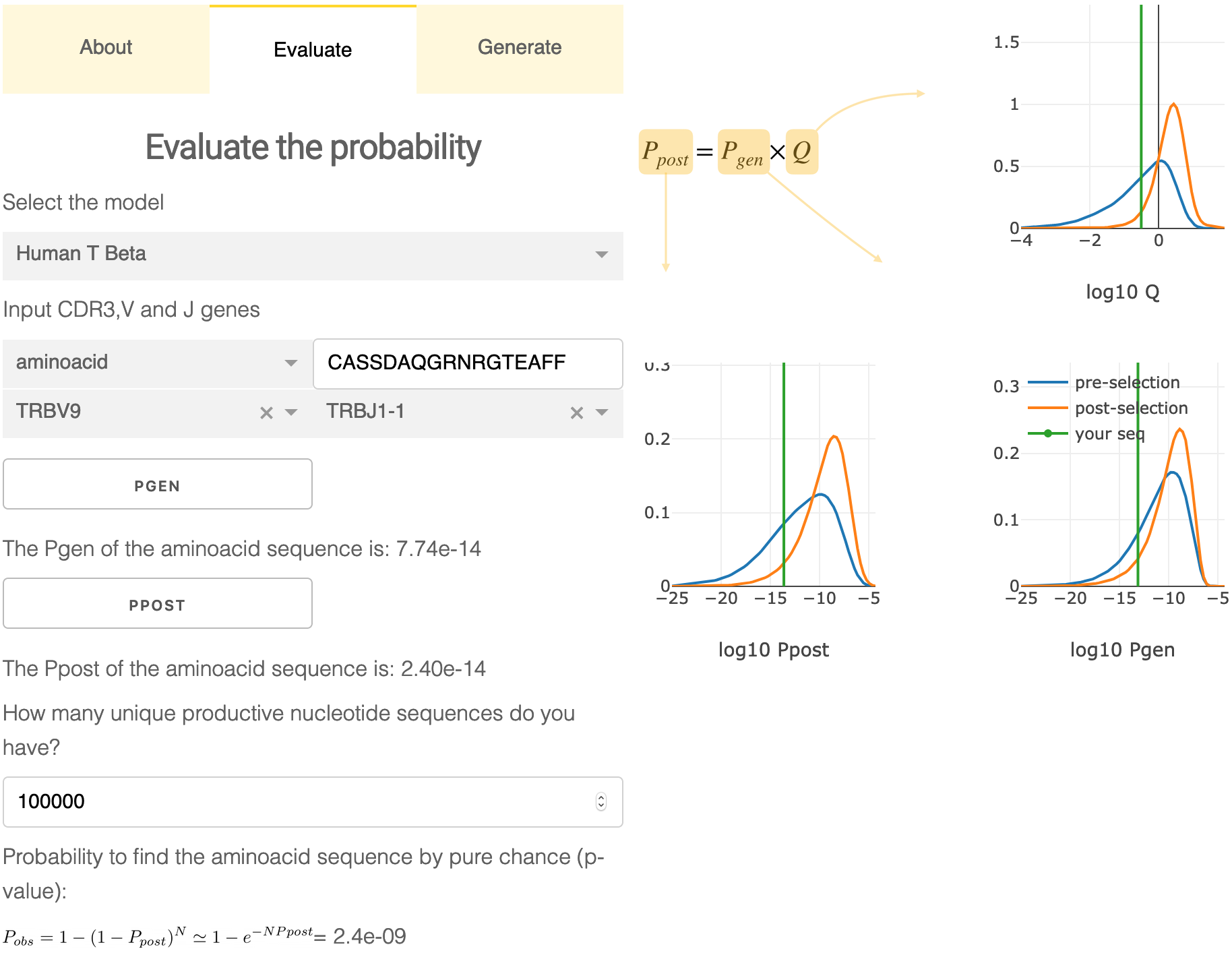}
\caption{
  \textbf{SOS web interface.} The user inputs a CDR3 sequence (amino acid or nucleotides) and V and J segments. The program outputs the generation probability $P_{\rm gen}$, the probability in the periphery $P_{\rm post}$, and evaluates a p-value corresponding to the probability of finding that sequence by chance in a repertoire of size $N$ (input by user). An additional tab allows for the generation of synthetic repertoires.
}
\label{fig}
\end{center}
\end{figure*}

After choosing the species and receptor chain in the ``Evaluate'' tab,
the user inputs a Complementary Determining Region 3 (CDR3), either as
a nucleotide or an amino acid sequence, and
optionally V and J germline genes from dropdown lists. The server outputs the generation probability ($P_{\rm gen}$,
conditioned on sequence productivity),
and the post selection probability ($P_{\rm post}$), as shown in
Fig.~\ref{fig} (left). When V and J are not specified, the program sums
over all possibilites for these segments to calculate the total
probability of the CDR3.

To help interpret the result and assess how the sequence of interest compares to others, $P_{\rm gen}$, $P_{\rm post}$, and the selection factor $Q=P_{\rm post}/P_{\rm gen}$ are plotted as green vertical lines on histograms of random sequences taken pre- (blue line) and post- (orange line) selection (Fig.~\ref{fig}, right). That feature only works when V and J and specified.
The tool also provides an estimation of the probability to observe the
sequence in a generic repertoire. The user
inputs the size $N$ of the sequenced repertoire (unique productive
nucleotide sequences), and the tool outputs the probability of
observing the sequence within a repertoire of that size, given by  $1-(1-P_{\rm post})^N$.

Using the ``Generate'' tab, the user can synthetize a specified number
of receptor sequences from $P_{\rm gen}$ or $P_{\rm post}$, after
choosing the species and chain type from dropdown lists. The file with
the generated sequences, composed of the CDR3 sequence (nucleotide and
amino-acid translation), V and J segments, is available for download
as a CSV file. The user may fix the seed of the random number
generator for reproducibility.

\section{Discussion}
\noindent The interface can be used by investigators to evaluate how
surprised one should be to find a given sequence in one or multiple repertoires. It
could help distinguish receptors with a specific function from chance detections.
The tool can also be used to evaluate the potential of certain
receptors (in particular antibodies, albeit in their unmutated version) for vaccination or therapeutic purposes.
The web interface is also  available on mobile phones without the plotting options.

\bigskip
{\bf Acknowledgments. }This work was partially supported by the European Research Council Proof of Concept Grant n. 824735

\bibliographystyle{pnas}

\begin{thebibliography}{10}

\bibitem{Marcou2018}
Marcou Q, Mora T, Walczak AM
\newblock (2018) {High-throughput immune repertoire analysis with IGoR}.
\newblock \emph{Nature Communications} 9:561.

\bibitem{Sethna2019}
Sethna Z, Elhanati Y, Callan CG, Walczak AM, Mora T
\newblock (2019) {OLGA: fast computation of generation probabilities of B- and
  T-cell receptor amino acid sequences and motifs}.
\newblock \emph{Bioinformatics} 35:2974--2981.

\bibitem{Sethna2020}
Sethna Z, {et~al.}
\newblock (2020) {Population variability in the generation and thymic selection
  of T-cell repertoires}.
\newblock \emph{arXiv:2001.02843} pp 1--17.

\bibitem{Murugan2012}
Murugan A, Mora T, Walczak AM, Callan CG
\newblock (2012) {Statistical inference of the generation probability of T-cell
  receptors from sequence repertoires}.
\newblock \emph{Proceedings of the National Academy of Sciences}
  109:16161--16166.

\bibitem{Ralph2016}
Ralph DK, Matsen FA
\newblock (2016) {Consistency of VDJ Rearrangement and Substitution Parameters
  Enables Accurate B Cell Receptor Sequence Annotation}.
\newblock \emph{PLoS Computational Biology} 12:1--25.

\bibitem{Elhanati2014}
Elhanati Y, Murugan A, Callan CG, Mora T, Walczak AM
\newblock (2014) {Quantifying selection in immune receptor repertoires}.
\newblock \emph{Proceedings of the National Academy of Sciences}
  111:9875--9880.

\bibitem{Georgiou2014}
Georgiou G, {et~al.}
\newblock (2014) {The promise and challenge of high-throughput sequencing of
  the antibody repertoire.}
\newblock \emph{Nature biotechnology} 32:158--68.

\bibitem{Heather2017}
Heather JM, Ismail M, Oakes T, Chain B
\newblock (2017) {High-throughput sequencing of the T-cell receptor repertoire:
  pitfalls and opportunities}.
\newblock \emph{Briefings in Bioinformatics} 19:554--565.

\bibitem{Minervina2019a}
Minervina A, Pogorelyy M, Mamedov I
\newblock (2019) {TCR and BCR repertoire profiling in adaptive immunity}.
\newblock \emph{Transplant International} pp 0--2.

\bibitem{Bradley2019}
Bradley P, Thomas PG
\newblock (2019) {Using T Cell Receptor Repertoires to Understand the
  Principles of Adaptive Immune Recognition}.
\newblock \emph{Annual Review of Immunology} 37:547--570.

\bibitem{Isacchini2020}
Isacchini G, {et~al.}
\newblock (2020) {On generative models of T-cell receptor sequences}.
\newblock \emph{arXiv:1911.12279}.

\end{thebibliography}

\end{document}